# Polarity effect on the heteroepitaxial growth of $B_xC$ on 4H-SiC by CVD

F. Cauwet[1a], Y. Benamra[1b], L. Auvray[1c], J. Andrieux [1d], G. Ferro[1e]

[1]Laboratoire des Multimatériaux et Interfaces, UMR CNRS 5615, Université de Lyon, 6 rue Victor Grignard, 69622 Villeurbanne, France

[a]francois.cauwet@univ-lyon1.fr, [b]yamina.benamra@etu.univ-lyon1.fr, [c]laurent.auvray@univ-lyon1.fr, [d]jerome.andrieux@univ-lyon1.fr, [e]gabriel.ferro@univ-lyon1.fr

**Abstract.** The chemical vapor deposition (CVD) growth of boron carbide ($B_xC$) layers on 4H-SiC, 4°off substrates was studied. Depending on the polarity of the substrate, different results were obtained. On Si face, the direct CVD growth at 1600°C under a mixture of $BCl_3+C_3H_8$ systematically led to polycrystalline $B_xC$ films, whatever the C/B ratio in the gas phase. On the C face, heteroepitaxial growth was obtained for C/B ratios = 12 or higher with a step bunched morphology. If a boridation step (10 min at 1200°C under $BCl_3$ flow) was used before the CVD growth, then heteroepitaxy was successful on both substrate polarities. To explain these results, a mechanism is proposed which involves the nature of the chemical bonds at the early stage of nucleation. It is suggested that a full B coverage of the SiC surface should favor the nucleation of the B-rich (0001) plane of $B_xC$, promoting thus the heteroepitaxial growth along this direction.

## Introduction

Boron carbide ($B_xC$) is a poorly known semiconductor, probably due to the difficulties in its crystalline elaboration under bulk [1] or thin film [2,3] form. Its estimated bandgap, ranging from 1.6 to 2.2 eV [4], associated to its thermal and chemical compatibility with SiC make it a good candidate for integration in 4H-SiC electronics, potentially targeting new devices (e.g. heterojunctions) and/or providing new properties (e.g. neutron detection) to the existing ones. According to the literature, $B_xC$ can be grown epitaxially on both polarities of 4H-SiC but with different procedures. On the C face, direct growth can lead to heteroepitaxy [5,6]. This is not the case for Si face for which a boridation step (treatment under $BCl_3$ at 1200C) is required to avoid polycrystalline growth [7]. The difference of growth mode between both polarities is still not well understood. Souqui et al suggested that epitaxial growth is promoted by the ability of the surface to provide C atoms [5] but it does not match with the successful heteroepitaxy on the Si face with a boridation step. Note that Souqui's et al. results were obtained by chemical vapor deposition (CVD) using a single-source precursor (triethyl-borane) which fixes the C/B ratio in the gas phase to 6. As a matter of fact, they obtained graphite co-deposition (inclusions) inside the $B_xC$ heteroepitaxial layers. One would probably need to adjust the C/B ratio in the gas phase to avoid such co-deposit, by using for instance different precursors for C and B atoms. This additional degree of freedom could also help both improving the layer quality and better understanding the nucleation mechanism. In addition, since $B_xC$ is a solid solution with $9 \leq [C] \leq 20$ at%, one could check for any effect of the C content in the gas phase on $B_xC$ stoichiometry. We thus investigate in this work the CVD growth of $B_xC$ on 4H-SiC 4°off with both Si and C polarities using separate precursors for C and B elements.

## Experimental

CVD growth was performed in a homemade vertical cold wall reactor working at atmospheric pressure with 16 slm of purified $H_2$ as carrier gas. High purity boron trichloride ($BCl_3$, 1% diluted in Ar) and propane ($C_3H_8$, 5% diluted in $H_2$) were used as boron and carbon precursors, respectively.

For each growth run, two pieces of 4H-SiC 4°off wafers were used, one with C polarity and another with Si polarity. They were cleaned in a methanol ultrasonic bath before their introduction into the reactor. The sample holder, a SiC-coated graphite susceptor, was induction-heated, and its temperature was measured and monitored via an optical pyrometer. The growth procedure involved heating under $H_2$ up to 1600°C. After 5 min of etching at this temperature, both precursors were introduced simultaneously for 1h growth. The C/B ratio in the gas phase was varied from 0.5 to 24 while keeping a constant $BCl_3$ flux of 2.5 sccm. For comparison purpose, some growth runs included a boridation step (10 min at 1200°C under 2.5 sccm of $BCl_3$) before the CVD growth at 1600°C under a C/B of 1 (see [7] for more details).

**Results and discussion**

The surface morphologies obtained at different C/B ratios are shown in Fig. 1. One can see that B-rich gas phase composition leads to isolated islands for both polarities (Fig.1a,f) while a fully covering layer is always obtained upon increasing this ratio.

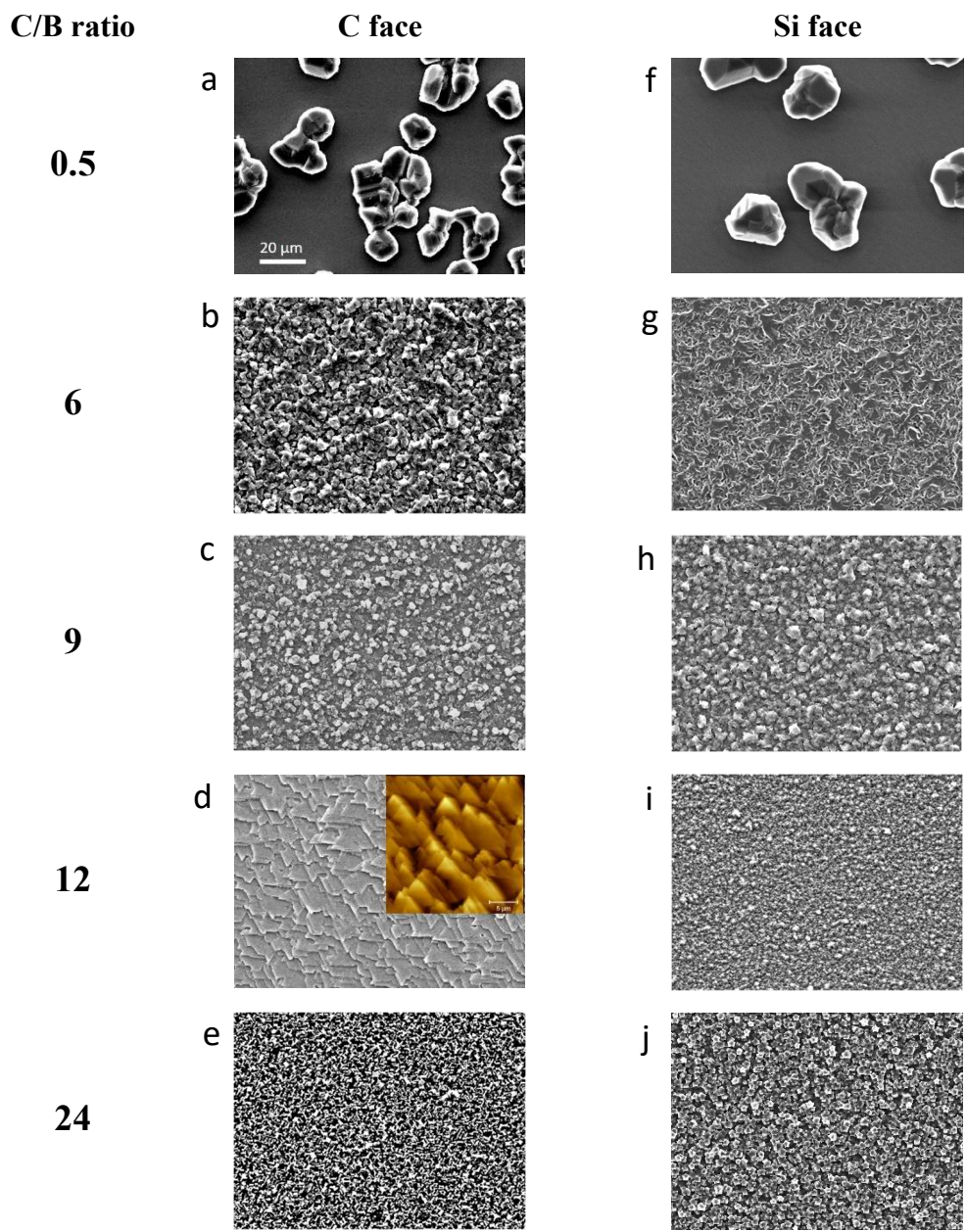

**Figure 1.** Surface morphology of the $B_xC$ layers grown at different C/B ratios, on Si- and C-face. The scale for all images is given in a). The inset in d) corresponds to a 25x25 µm$^2$ AFM scan of the same sample

Very rough and disordered surfaces are systematically found for deposition on Si face substrate though the fine microstructure changes depending on the C/B ratio (Fig. 1f-j). This is more or less similar on C face but with different microstructures (Fig. 1a-e). Interestingly, the sample grown on C face with CB=12 displays some oriented features, looking like a jagged step bunched morphology (Fig. 1d). This is better seen in the atomic force microscopy (AFM) inset image.

X-ray diffraction (XRD) analyses on these samples gave a different insight on this set of samples (Fig. 2). As can be seen, the crystalline orientation(s) of the $B_xC$ layers are significantly influenced by these two parameters. First of all, except for the case of C/B = 0.5, the layers grown on Si face do not exhibit the same diffracting planes in the XRD patterns as compared to C face ones: $(10\bar{1}4)$, $(02\bar{2}1)$ and $(20\bar{2}5)$ for Si face and $(0003)$, $(01\bar{1}2)$ and $(02\bar{2}4)$ for C face. While this does not change significantly when varying the C/B ratio on Si-face, the C-face behaves differently with the predominance of the [0003] peak for C/B = 9 and above. For C/B = 12, the only XRD peak is attributed to [0003] suggesting a preferred (epitaxial-like) out-of-plane orientation of the $B_xC$ layer, in correlation with the step-bunched morphology for this sample. No peak related to graphite phase is detected on this sample while, when the C/B ratio is increased up to 24, an intense additional peak appears at 2θ≈26°, attributed to 2H-graphite (0002), together with less intense C-related peaks. Note that this graphite phase does not affect the orientation of the $B_xC$ layer which remains preferentially oriented along the [0001] direction.

Interestingly, all the XRD peaks of the $B_xC$ layers grown at C/B = 0.5 are slightly shifted toward lower 2θ values as compared to the other layers grown at higher C/B ratios. This effect is better seen for C face samples due to the availability of several common peaks identically attributed for C/B = 0.5 and above. Such low angle shift could be related to a decrease of the C content inside $B_xC$ since it was shown to decrease the lattice parameters of the material [8]. For C/B ≥ 6, the $B_xC$ peaks do not shift anymore, suggesting that the corresponding films should have roughly the same composition. Assuming that the $B_xC$ layers grown with graphite co-deposit have the highest possible C content ($B_4C$ stoichiometry), then all the layers grown with C/B ≥ 6 should also have this stoichiometry. Rutherford Backscattering spectroscopy analyses are currently underway to confirm this point.

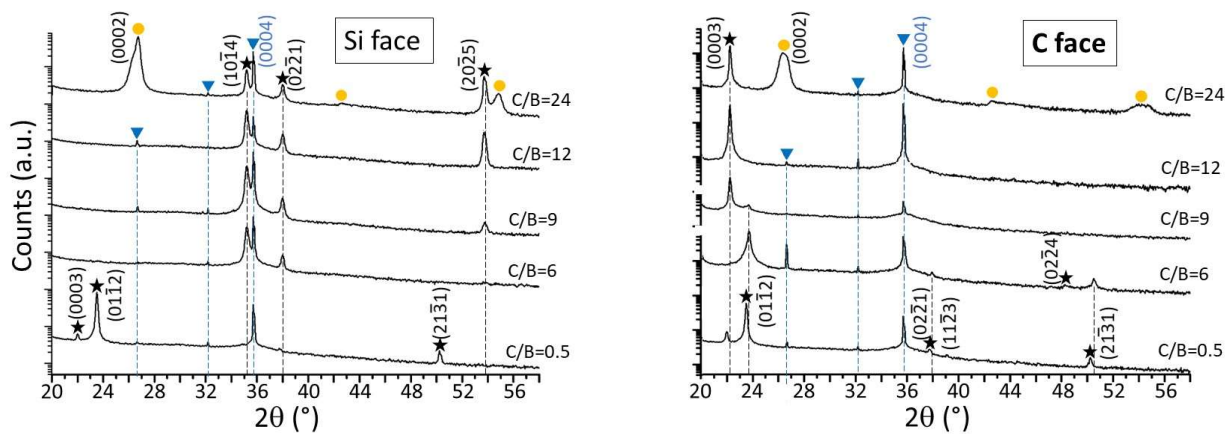

**Figure 2.** XRD patterns recorded on the $B_xC$ layers grown on Si- (left) and C-face (right) 4H-SiC substrates with various C/B ratios. (▼ = 4H-SiC; ★ = $B_xC$; ● = Graphite)

We have shown in ref [7] that using a boridation step was efficient to promote the heteroepitaxial growth of $B_xC$ on Si face 4H-SiC substrate. This procedure was thus tested on C face substrate to check for any difference with the Si face case. The results obtained on both polarities with the modified growth procedure are shown in Fig. 3. One can see that both samples display oriented features at their surface. The XRD patterns show only one $B_xC$ peak attributed to the (0003) planes. From these results, we can assume that both layers are heteroepitaxial. As a matter of fact, the boridation step works on both polarities to favour $B_xC$ [0001] out-of-plane orientation. This result does not match with the hypothesis proposed in [5] which considers the ability of the surface to provide C atoms. The mechanism behind the polar-dependent heteroepitaxy observed here is probably more complex. One can get complementary information by considering the crystal structure of $B_4C$ (Fig.4). It is composed of $B_{12}$ icosahedra partly connected with triatomic chains containing C atoms. One can see that the (0001) plane is composed of a dense packing of $B_{12}$ icosahedra; it is thus a boron-rich plane. Let us now see how $B_xC$ can nucleate on 4H-SiC considering the separate cases of direct CVD and boridation+CVD. This is schematically summarized on Fig. 5. Starting with direct CVD, the C and B atoms brought by the gas phase can theoretically stick to the 4H-SiC surface by creating C-Si and B-Si bonds on Si face and C-C and B-C bonds on C face.

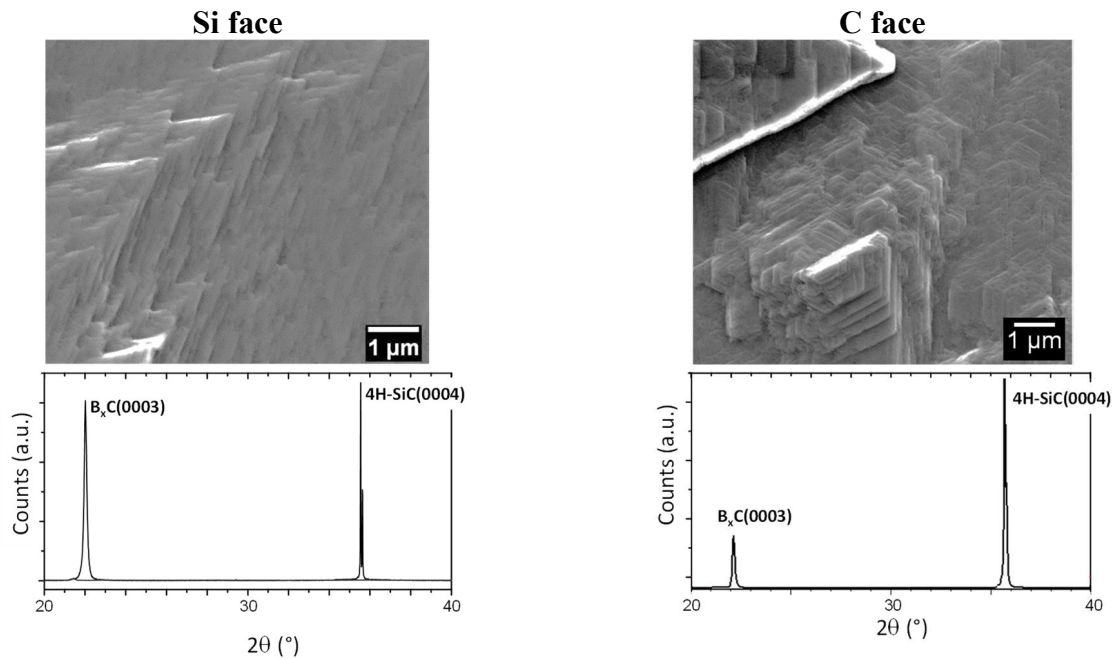

**Figure 3.** SEM images and XRD patterns obtained on $B_xC$ layers grown on Si- (left) and C-face (right) 4H-SiC substrates using the modified procedure including the boridation step.

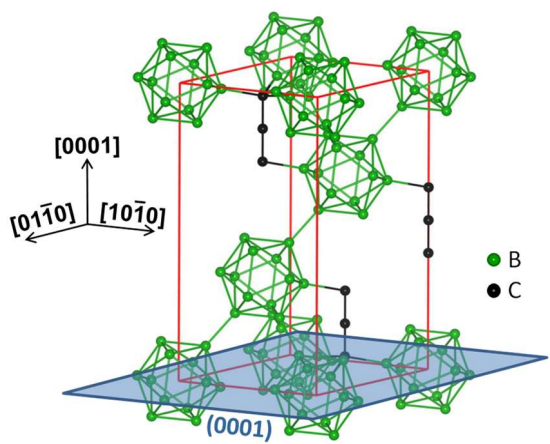

**Figure 4**. Schematic crystallographic structure of $B_4C$ showing the hexagonal unit cell.

However, we believe that C-C bonds can hardly form due to $H_2$ reducing effect at such high temperature of 1600°C and under high propane dilution in $H_2$ (for C/B=12 this dilution is 1600). For instance, that is the reason why graphene is mostly grown (by sublimation or CVD) under Ar or Ar+$H_2$ mixture as carrier gas [9,10]. Also, it was shown that a latency period (which can be as long as several minutes) exists before starting graphene nucleation [11]. Therefore, on C face, B atoms can form the kinetically most stable bond with the C uppermost atoms of the SiC surface. Note that higher propane partial pressures (C/B=24) lead to graphite inclusions inside the layers, without affection though the epitaxial nature of the $B_4C$ layer which means that the nucleation step is probably not affected. On Si face, both Si-B and Si-C bonds are stable though with different energies (289 and 435 kJ.mol$^{-1}$ respectively for Si-B and Si-C bonds [12]). The successful heteroepitaxial growth on C face by direct CVD could thus be explained by the preferred formation of C-B bonds at the interface with the substrate leading to a B-rich nucleation layer. This should favor the nucleation of the (0001) plane since it requires the highest density of B atoms within the $B_4C$ crystal structure. On Si face, the possibility of Si-C bonds formation (along with the Si-B ones) should lead to the nucleation of other planes than (0001). Several possibilities exist so that a polycrystalline layer is formed.

In the case of the boridation+CVD, the SiC surface is only in contact with a B-containing gas in the first instants of boridation so that it should saturate the SiC surface with Si-B and C-B bonds on Si and C polarity respectively. Again, this should favour the heteroepitaxial growth of $B_xC$ but this time on both polarities. Note that the CVD step after boridation was performed with a C/B ratio as low as 1, without affecting the heteroepitaxial orientation of the layer. In the case of direct CVD, we have shown earlier that heteroepitaxy was not possible with such a low C/B value. The reason for this is still unclear and will require further experiments to better understand the mechanism in play.

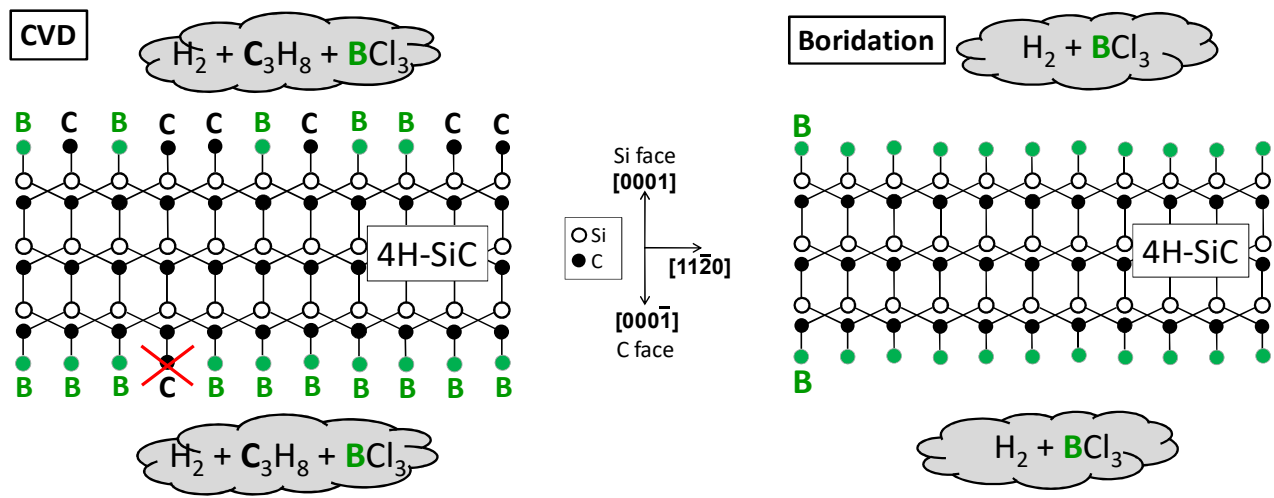

**Figure 5**. Schematic representation of the bonding configuration at $B_xC$ nucleation stage for both cases of direct CVD (left) and boridation (right). The red cross on C face for CVD case illustrates the fact that C-C bonds are not favoured on C face, as discussed in the text.

**Summary**

We have studied the heteroepitaxial growth of $B_xC$ layers on C- and Si-polar 4H-SiC 4°off substrates by CVD. We have confirmed that C face is more suitable for successful heteroepitaxy than Si face, thanks to the preferential formation of C-B bonds at the early stage of nucleation. This should lead to the initial formation of the B-rich (0001) plane of $B_xC$ which should then promote the heteroepitaxial growth along the [0001] direction. Heteroepitaxy on Si face substrate is also possible by inserting a boridation step which allows also nucleating with B-rich (0001) plane.